# Effects of Side Groups on Kinetics of Charge Carrier Recombination in Dye Molecule-Doped Multilayer Organic Light-Emitting Diodes

Shengwei Shi[1,2,*], Feng Gao[2], Zhengyi Sun[2], Yiqiang Zhan[3], Mats Fahlman[2], Dongge Ma[1]

[1]State Key Laboratory of Polymer Chemistry and Physics, Changchun Institute of Applied Chemistry, Chinese Academy of Sciences, 130022, Changchun, P. R. China
*Corresponding author, email: shesh@ifm.liu.se
[2]Department of Physics, Chemistry and Biology, Linköping University, S-58183 Linköping, Sweden
[3]State Key Laboratory of ASIC and System, Department of Microelectronics, SIST, Fudan University, 200433, Shanghai, China

**Abstract:** The kinetics of the charge carrier recombination in dye molecule-doped multilayer organic light-emitting diodes (OLEDs) was quantified by transient electroluminescence (EL). Three sets of dye molecules, such as derivatives of naphthalimide and stilbene, were used as dopants in light-emission layer. Although the devices show almost the same EL spectra for each set of molecules, they show very different EL efficiency. The difference in EL efficiency was attributed to the difference in charge carrier recombination, as revealed by transient EL. The recombination coefficient ($\gamma$) was determined from the long-time component of the temporal decay of the EL intensity after a rectangular voltage pulse was turned off. It was found that $\gamma$ and EL efficiency were both strongly dependent on the molecular structures of the dopants, and the donor groups and $\pi$-conjugated structure guaranteed high $\gamma$ and EL efficiency in OLEDs.

## 1. Introduction

Organic light-emitting diodes (OLEDs) have been extensively studied for their promising applications in flexible displays [1] and solid state lighting.[2,3] High performance organic materials, new device structures and advanced processing technologies are developed to advance the development of the OLED community.[4-6] So far, OLEDs are the most successful application of organic semiconductors. It is well acknowledged that electrical processes in OLEDs include three key steps, i.e. charge injection, charge transport, and charge recombination. All of these three processes play very important roles in device performance.[7-11]

Transient electroluminescence (EL) (time resolved EL) has been widely used to investigate the electrical processes in OLEDs, enabling us to analyze charge transport and recombination in OLEDs.[12-22] When a square voltage pulse is applied over the device, the EL starts to rise with a time delay, which has been interpreted as the transit time of the majority charge carriers. As the RC time constant of the setup is well below the delay time, it is uklikely that the RC constant is responsible for the delay time. The interpretation of the response time of the device in terms of transit time gives the mobility value.[12-18] When the square voltage pulse is removed, the EL starts to decay, and it gradually decreases to zero EL. This decay time is closely related with bimolecular recombination of electrons and holes. The recombination coefficient ($\gamma$) then can be obtained from the decay part in transient EL,[19-21] which is considered to be very important to understand the operating mechanism in OLEDs. However, most mechanistic investigations of OLEDs from transient EL are focused on the simple strucutre of electron-tranpsort layer/emissive layer/hole-transport layer;[12-21] there are less device phyiscs studies on the multilayer structure,[22] which is adopted in practical aplications. In addition, although many highly efficient molecules are developed in OLEDs, with different side groups to adjust the solubility,[23] suppress the crystallization at ITO/organic interface,[24] reduce concentration quenching in phosphorescence,[25] or inhibit the aggregation light-emission,[26] it is not clear how these side goups affect the device efficiency from the understanding of device physics.

In this communication, we employed transient EL to investigate charge recombination in multilayer OLEDs. We used dye-doped CBP to work as the emissive layer. By keeping the backbone the same and changing the side groups, we were able to quantify the effect of the side groups on charge recombination, and hence the device performance. The recombination coefficient ($\gamma$) of charge carriers are obtained from the decay part of transient EL through elaborate design of device structure. Our results are in good agreement with the predictions of the Langevin model of charge carrier recombination assuming the process controlled by charge carrier diffusion.[27] The analysis of charge carrier recombination kinetics clearly demonstrates that the donor groups and π-conjugated structure of the molecule assure high EL efficiency with the same EL spectra. Based on the results of device performance and transient EL, we discuss the relationship between $\gamma$ and molecular structure, which may guide the molecule design by modification of side groups to realize highly efficient OLEDs.

**2. Results and discussion**

The dye molecule dopants used in this experiment are derivatives of naphthalimide and stilbene, which have been reported to be very promising molecules in OLEDs. They have very good photochemical and thermal stability, and their molecule structures can be easily modified to obtain color adjustable emission.[28-33] We use three different dye molecules (**Figure 1**) in the experiment. Considering that consistent results can be drawn from these three different molecules, we focus on the discussions on Molecule 1 in the main text and leave Molecules 2 and 3 in the Supplementary Information. **Table 1** summarises the physical properties of Molecule 1 with different side groups.

The device structure is shown in **Figure 2** and the related energy levels are collected from the references.[34-36] As shown in **Figure 3a**, four devices have almost the same EL spectra peaked at around 500 nm. This peak is significantly different from the peak at ~ 550 nm from pure $Alq_3$,[37] meaning that the carrier recombination is mainly confined in the molecule doping layer. At the same time, we notice that the four molecules with different side groups give very different device performance (Figure 3b). **Table 2** lists the performance for all four devices. Device 1a has the highest current efficiency of 3.7 cd/A with luminance of 3640 $cd/m^2$, while Device 1d gives the lowest efficiency (less than 1 cd/A) with luminance about 760 $cd/m^2$. The performance of Devices 1b and 1c lies between that of 1a and 1d.

In order to understand the operation mechanisms of the devices as well as the physics behind the significant difference between these four devices, we examined their J-V curves. As shown in **Figure 4a**, all the the J-V characteristics follow power-law dependence, which has been proposed to be charge-trap limited transport.[38,39] Assuming an exponential trap energy distribution, the J-V relationship can be described by $J \sim V^{m+1}$, where $m = T_t/T$ with T being the absolute temperature and $T_t$ being the characteristic temperature of the trap distribution. The value of the exponent m is related with trap concentrations and mobility, and higher value of m means deeper traps in the devices. For all the four devices, they share an exponent value of around 9, indicating that there are similar trap distributions in the four devices. Considering the dye dopants have similar molecular structures, it is reasonable that the four devices have similar trap distributions. The fact that the current is dominated by traps also indicates that J-V curves are determined by the the bulk properties rather than by contact effects.[19]

Further information regarding the device operation is revealed from the the luminance-current density (L-J) curves. As shown in Figure 4b, all the L-J curves have a linear relationship in log-log plot with a slope of 1 at high electric fields, meaning that the devices work in the volume-controlled EL mode,[19] consistent with our conclusion based on the J-V curves. Both the power-law dependence of J-V and the linear relation of L-J confirm the Langevin bimolecular recombination in the devices.[27] Holes injected from the ITO are transported through the hole transport layer NPB to the doping emission layer, and their transport to Alq$_3$ side is blocked by the BCP layer because of high injection barrier at CBP/BCP interface. Therefore, holes are confined in the doped emission layer and have to wait for the electrons to arrive. Electrons injected from LiF/Al cathode are transported through the electron transport layer Alq$_3$ and the hole blocking layer BCP. Electrons finally transport to the doped emission layer to recombine with the waiting holes, producing light emission from the dye molecule. The low mobilities of NPB and Alq$_3$ make the effective transit time in the device very long, resulting in high recombination probability.[12,19] As a result, the devices operate in the volume-limited mode.

For volume-limited current, with bimolecular recombination in the narrow zone of the emission layer to be the main channel for carriers decay, the free carrier kinetics can be described by the following equations:[19]

$$\frac{dn_h}{dt} \cong \frac{j_h}{eL_h} - \gamma n^2 \qquad (1)$$

$$\frac{dn_e}{dt} \cong \frac{j_e}{eL_e} - \gamma n^2 \qquad (2)$$

where n, j and L are the charge carrier concentration, current density and the thickness of transport layer, respectively. $\gamma$ is the bimolecular recombination coefficient. The subscripts h and e represent hole and electron. Here, "electron transport" layer can be approximately considered to include Alq$_3$ and BCP since there is almost no barrier for electron injection at the interface; "hole transport" layer includes NPB and CBP because of small barrier for hole injection and its higher mobility. As a result, the transport layer has the same thickness of 60 nm for electrons and holes. Because $j_h = j_e = j$, the initial concentration of charge carriers is $n_0 = (j/e\gamma L)^{1/2}$ under the steady-state conditions ($L_h = L_e = L$).

Although the hole mobility in NPB is larger than the electron mobility in Alq$_3$, there is a high barrier at CBP/BCP interface for hole injection into Alq$_3$ layer. As a result, the excess holes will accumulate in doped emission layer near CBP/BCP interface. For electrons, it is easier for them to inject through BCP to the emission layer. Therefore the light-emission is dominated by the recombination in doped layer rather than in Alq3, as confirmed by the spectra results. In addition, because of the relatively large value of the coulombic capture radius, the recombination time is substantially shorter than the trapping time. This means that the injected electrons and holes in the recombination zone will recombine rapidly, and hence no excess charge carriers exist. Consequently holes as well as electrons in the recombination zone may be considered to be free and approximately equal each other in concentration ($n_h \cong n_e = n$). The charge decay after the voltage pulse is turned off can then be simplified as

$$\frac{dn}{dt} \cong -\gamma n^2 \tag{3}$$

Thus,

$$\frac{1}{n} = \frac{1}{n_0} + \gamma t \tag{4}$$

Taking into account the EL yield,

$$\Phi_{EL} = \varphi_{PL} P_s \gamma [n(t)]^2 \tag{5}$$

where $\varphi_{PL}$ is the fluorescent quantum efficiency, $P_s$ is the function that a singlet rather than a triplet is generated. The EL decay can then be expressed as

$$\frac{1}{\sqrt{\Phi_{EL}(t)}} = \frac{1}{\sqrt{\varphi_{PL} P_s n_0^2}} + \sqrt{\frac{\gamma}{\varphi_{PL} P_s}} t \tag{6}$$

This equation clearly indicates that the reciprocal of the square root of the decay EL intensity is in linear relashionship with the time scale. From the ratio of slope $S = (\gamma/\varphi_{PL} P_s)^{1/2}$ to intercept $A = (\varphi_{PL} P_s n_0^2)^{-1/2}$, the electron-hole recombination coefficient then can be calculated as:

$$\gamma = \frac{(S/A)^2 eL}{j} \tag{7}$$

Equation 7 enables us to investigate the charge recombination immediately after switching off the bias. Therefore, we employed transient EL to investigate charge recombination in the devices, aiming to further understand the difference between these four devices.

Figure 5a shows the relationship of transient EL intensity-time in dye molecule-doped OLEDs at the same equilibrium current of 170 mA/cm$^2$ and L= 60 nm (refer to Equation 7), and it shows different EL decay for devices with different doping molecules, providing useful information on the carrier recombination in different devices. In order to quantify the charge recombination, we further plot the transient EL decay $(\varphi_{EL})^{-1/2}$ versus time in Figure 5b, where the inset shows the $\varphi_{EL}(t)$ decay curves. Here t=0 corresponds to the voltage fall of the pulse. Perfect straight lines can be fitted to the experimental data withing experimental error range. The values of slopes (S) and intercepts (A) are obtained from the linear relationship, and then the charge carrier recombination coefficient γ can be calculated (refer to Equation 7). The values are summarised in Table 2. The device with Molecule 1a doping has the highest γ of 5.8×10$^{-12}$ cm$^3$/s, while the device with Molecule 1d doping has the lowest γ. As well known, high γ means that highly efficient recombination of holes and electrons in the devices. Therefore the fitted γ values are consistent with the device performance, with the device with Molecule 1a showing the highest efficiency. Considering that the only difference between different molecules lies in the side groups, it is reasonable to conclude that the difference in EL efficiency may result from the side groups. Donor group (e.g. in Molecule 1a) is considered to form strong push-pull electron structure, and hence the electron can easily transfer to the aromatic ring to form expanded π-conjugated structure, enhancing fluorescence emission from the molecule backbone.[28,29] Strong acceptor group (e.g. in Molecule 1d) will decrease the fluorescence emission as it reduces the electron density in the system. Indeed, in our experiments, donor groups assure higher EL efficiency for the molecule-doping multilayer OLEDs, and the higher γ values are consistent with the device results.

We also extended our measurements to other molecules to generalise our observasatioins. For Molecule 2, we obtained similar results and conclusions as Molecule 1 (Figures s1 and s2), in which the donor groups, such as –$CH_3$, guarante higher γ and EL efficiency, while the acceptor groups, such as –F and –$CF_3$, decrease the γ and EL efficiency. Molecule 3, though a little bit complicated, also shows results consistent with Molecules 1and 2. $R_1$ group shows stronger donor in 3a and 3b (t-Bu) than in 3c (-H). In addition, R2 group, -$COOCH_3$ (3a) and p-$C_6H_6$ (3b) can bring bigger π-conjugated structure than 2,5-$OCH_3$ (3c) in the backbone. Therefore, more delocalized electrons can be excited, and then the devices with Molecules 3a and 3b show higher γ and EL efficiency compared with Molecule 3c (Figure s3 and s4).

**3. Conclusion**

Multilayer OLEDs was designed and fabricated based on molecule-doping technology, in which derivatives of naphthalimide and stilbene with different side groups were used as the dopants in CBP host as the light-emission layer. All devices for each set of molecules showed almost the same EL spectra, but very different efficiency. Transient EL, by which the kinetics of the charge carrier recombination were investigated, was used to understand the physics behind the different EL efficiency.The coefficient (γ) of electron-hole recombination was determined from the long-time component of the temporal decay of the EL intensity after a rectangular voltage pulse was turned off. It was found that γ and EL efficiency were both strongly dependent on the side groups of the dopants. Donor structures in the side groups guarantee much higher γ and EL efficiency than the acceptor ones. Our results provide a promising guide for the structure design for molecular materials in highly efficient OLEDs. Chemists can slightly change the side groups to maintain the desired emission spectra while significantly enhance the device efficiency. In addition, this report may provide useful hints on how to design molecule structure to reduce charge carrier recombination, which is considered to be the main cause of efficiency loss in organic solar cells.[40-42]

## 4. Experimental Section

4.1. Substrate and active layers

The device had a molecule-doped multilayer structure with pre-cleaned indium–tin-oxide (ITO) as the anode and aluminium (Al) as the cathode. N,N´-Bis(naphthalen-1-yl)-N,N´-bis(phenyl)benzidine (NPB) and tris(8-hydroxyquinoline) aluminium (Alq$_3$) were used as hole and electron transport layer, respectively. 4,4´-Bis(N-carbazolyl)-1,1´-biphenyl (CBP) was employed as the host, and Bathocuproine (BCP) worked as the hole blocking layer. LiF was deposited as the buffer layer. Dye molecules with different side groups were used as dopants (Figure 1).

4.2. Device fabrication

The devices were fabricated with the structure: ITO/NPB (40 nm)/CBP:X (1 %)(20 nm)/BCP (20 nm)/Alq$_3$ (40 nm)/LiF (1 nm)/Al (200 nm), in which X indicated the dye molecules used in the experiments (Figure 2). All layers were prepared in a high-vacuum chamber with the pressure less than $5\times10^{-4}$ Pa by thermal evaporation without breaking the vacuum. All organic films except dye molecules were deposited at the rate of 0.1~0.3 nm/s, and the doping (1%) was realized by controlling the rate ratio between the host and dye molecules. LiF and Al were deposited at a rate of 0.01 nm/s and 0.8~1.0 nm/s, respectively. The rate is determined by a calibrated quartz microbalance, and the active area is $3 \times 3$ mm$^2$.

4.2. Device characterization

Current density-Luminance-Voltage (J-L-V) characteristics were measured using a Keithley sourcemeter unit (Keithley 2400 and Keithley 2000) with a calibrated silicon photodiode. The EL spectra were measured by a JY SPEX CCD3000 spectrometer. The UV-vis and PL spectra were measured using a Perkin-Elmer Lambda 35 UV-vis spectrometer and a Perkin-Elmer LS 50B spectrofluorometer, respectively. For the transient EL, an Agilent 8114A 100 V/2A programmable pulse generator was used to apply rectangular voltage pulse to the devices. The repetition rate of the pulse was 1 kHz, and the pulse length was 20μs. The time-dependent EL signals were detected by the 50 Ω input resistance of a digital oscilloscope (Agilent Model 54825A, 500 MHz/2 Gs/s) together with a photomultiplier (time resolution ≅0.65 ns) located directly on top of the emitting devices. All measurements were carried out at room temperature under ambient conditions.

**Apendix A. Supplementary material**

In this supplementary material, experimental results for dye molecule-doped multilayer OLEDs with other molecular systems (2 and 3) are listed as below by use of steady state and transient EL characterization to support the conclusion and discussion in the main text.

**Acknowledgements**

We thank Prof. Longhe Xu from Dalian University of Technology to provide the molecules. Dr. Ma thanks the financial support of Hundreds Talents Program, Chinese Academy of Sciences and the National Science Fund for Distinguished Young Scholars of China (50325312).  F.G. acknowledges the financial support of the European Commission under a Marie Curie Intra-European Fellowship for Career Development.

**Reference:**

[1]. G. Gustafsson, Y. Cao, G.M. Treacy, F. Klavetter, N. Colaneri, A.J. Heeger, Flexible light-emitting diodes made from soluble conducting polymers, Nature 357 (1992) 477-479.
[2]. B.W. D'Andrade, S.R. Forrest, White organic light-emitting devices for solid-state lighting, Adv. Mater. 16 (2004) 1585-1595.
[3]. S. Reineke, F. Lindner, G. Schwartz, N. Seidler, K. Walzer, B. Lussem, K. Leo, White organic light-emitting diodes with fluorescent tube efficiency, Nature 459 (2009) 234-238.
[4]. L. Xiao, Z. Chen, B. Qu, J. Luo, S. Kong, Q. Gong, J. Kido, Recent progresses on materials for electrophosphorescent organic light-emitting devices, Adv. Mater. 23 (2011) 926-952.
[5]. M.C. Gather, A. Köhnen, K. Meerholz, White organic light-emitting diodes, Adv. Mater. 23 (2011) 233-248.
[6]. G. Schwartz, S. Reineke, T.C. Rosenow, K. Walzer, K. Leo, Triplet harvesting in hybrid white organic light-emitting diodes, Adv. Funct. Mater. 19 (2009) 1319-1333.
[7]. W. Brutting, S. Berleb, A.G. Muckl, Device physics of organic light-emitting diodes based on molecular materials, Org. Electron. 2 (2001) 1-36.
[8]. M.A. Baldo, S.R. Forrest, Interface-limited injection in amorphous organic semiconductors, Phys. Rev. B. 64 (2001) 085201.
[9]. J.C. Scott, Metal–organic interface and charge injection in organic electronic devices, J. Vac. Sci. Technol. A 21 (2003) 521-531.
[10]. G.G. Malliaras, J.C. Scott, The roles of injection and mobility in organic light emitting diodes, J. Appl. Phys. 83 (1998) 5399-5403.

[11]. Y. Shirota, H. Kageyama, Charge carrier transporting molecular materials and their applications in devices, Chem. Rev. 107 (2007) 953-1010.

[12]. T.C. Wong, J. Kovac, C.S. Lee, L.S. Hung, S.T. Lee, Transient electroluminescence measurements on electron-mobility of N-arylbenzimidazoles, Chem. Phys. Lett. 334 (2001) 61-64.

[13]. S. Barth, P. Muller, H. Reil, P.F. Seidler, W. Rieß, H. Vestweber, H. Bassler, Electron mobility in tris(8-hydroxy-quinoline)aluminum thin films determined via transient electroluminescence from single- and multilayer organic light-emitting diodes, J. Appl. Phys. 89 (2001) 3711-3719.

[14]. H. Chayet, R. Pogreb, D. Davidov, Transient uv electroluminescence from poly(p-phenylenevinylene) conjugated polymer induced by strong voltage pulses, Phys. Rev. B 56 (1997) R12702.

[15]. P.W.M. Blom, M.C.J.M. Vissenberg, Dispersive hole transport in poly( p-phenylene vinylene), Phys. Rev. Lett. 80 (1998) 3819-3822.

[16]. D.J. Pinner, R.H. Friend, N. Tessler, Transient electroluminescence of polymer light emitting diodes using electrical pulses, J. Appl. Phys. 86 (1999) 5116-5130.

[17]. Y.H. Tak, J. Pommerehme, H. Vestweber, R. Sander, H. Bassler, H.H. Horhold, Pulsed electroluminescence from organic bilayer light emitting diodes, Appl. Phys. Lett. 69 (1996) 1291-1293.

[18]. J. Chen, D. Ma, Effect of dye concentration on the charge carrier transport in molecularly doped organic light-emitting diodes, J. Appl. Phys. 95 (2004) 5778-5781.

[19]. J. Kalinowski, N. Camaioni, P. Di Marco, V. Tattori, A. Martelli, Kinetics of charge carrier recombination in organic light-emitting diodes, Appl. Phys. Lett. 72 (1998) 513-515.

[20]. J. Chen, D. Ma, Effect of dye-doped concentration on the charge carrier recombination in molecularly doped organic light-emitting devices, J. Phys. D: Appl. Phys. 39 (2006) 2044-2047.

[21]. J. Chen, D. Ma, Y. Liu, Y. Wang, Studies of kinetics of charge carrier recombination in organic light-emitting diodes based on beryllium complexes by transient electroluminescence, J. Phys. D: Appl. Phys. 38 (2005) 3366-3370.

[22]. M.A. Baldo, S.R. Forrest, Transient analysis of organic electrophosphorescence: I. Transient analysis of triplet energy transfer, Phys. Rev. B. 62 (2000) 10958-10966.

[23]. J.H. Park, C. Yun, M.H. Park, Y. Do, S. Yoo, M.H. Lee, Vinyl-type polynorbornenes with triarylamine side groups: a new class of soluble hole-transporting materials for OLEDs, Macromolecules 42 (2009) 6840-6843.

[24]. S.E. Shaheen, G.E. Jabbour, B. Kippelen, N. Peyghambarian, J.D. Anderson, S.R. Marder, N.R. Armstrong, E. Bellmann, R.H. Grubbs, Organic light-emitting diode with 20 lm/W efficiency using a triphenyldiamine side-group polymer as the hole transport layer, Appl. Phys. Lett. 74 (1999) 3212-3214.

[25]. C. Rothe, C.J. Chiang, V. Jankus, K. Abdullah, X. Zeng, R. Jitchati, A.S. Batsanov, M.R. Bryce, A.P. Monkman, Ionic Iridium(III) complexes with bulky side groups for use in light emitting cells: reduction of concentration quenching, Adv. Funct. Mater. 19 (2009) 2038-2044.

[26]. S. Setayesh, A.C. Grimsdale, T. Weil, V. Enkelmann, K. Mullen, F. Meghdadi, E.J.W. List, G. Leising, Polyfluorenes with polyphenylene dendron side chains: toward non-aggregating, light-emitting polymers, J. Am. Chem. Soc. 123 (2001) 946-953.

[27]. M. Pope, C.E. Swenberg, Electronic Processes in Organic Crystals, Oxford University Press, Oxford 1982.

[28]. J.X. Yang, X.L. Wang, X.M. Wang, L.H. Xu, The synthesis and spectral properties of novel 4-phenylacetylene-1,8-naphthalimide derivatives, Dyes Pigm. 66 (2005) 83-87.

[29]. J.X. Yang, X.L. Wang, S. Tong, L.H. Xu, Studies on the synthesis and spectral

properties of novel 4-benzofuranyl-1,8-naphthalimide derivatives, Dyes Pigm. 67 (2005) 27-33.

[30]. P. Wang, Z. Xie, S. Tong, O. Wong, C.S. Lee, N. Wong, L. Hung, S.T. Lee, A novel neutral red derivative for applications in high-performance red-emitting electroluminescent devices, Chem. Mater. 15 (2003) 1913-1917.

[31]. W. Zhu, L. Fan, R. Yao, F. Wu, H. Tian, Naphthalimide incorporating oxadiazole: potential electroluminescent materials with high electron affinity, Synt. Met. 137 (2003) 1129-1130.

[32]. A. Kukhta, E. Kolesnik, M. Taoubi, D. Drozdova, N. Prokopchuk, Polynaphthalimide is a new polymer for organic electroluminescence devices, Synt. Met. 119 (2001) 129-130.

[33]. I. Fuks-Janczarek, I.V. Kityk, R. Miedzinski, E. Gondek, J. Ebothe, L. Nzoghe-Mendome, A. Danel, Push-pull benzoxazole based stilbenes as new promising electrooptics materials, J. Mater. Sci. Mater. Electron 18 (2007) 519-526.

[34]. S. Shi, D. Ma, A pentacene-doped hole injection layer for organic light-emitting diodes, Semicond. Sci. Technol. 20 (2005) 1213-1216.

[35]. M.A. Baldo, S. Lamansky, P.E. Burrows, M.E. Thompson, S.R. Forrest, Very high-efficiency green organic light-emitting devices based on electrophosphorescence, Appl. Phys. Lett. 75 (1999) 4-6.

[36]. T. Zhang, Y. Liang, J. Cheng, J. Li, A CBP derivative as bipolar host for performance enhancement in phosphorescent organic light-emitting diodes, J. Mater. Chem. C 1 (2013) 757-764.

[37]. C.W. Tang, S.A. VanSlyke, Organic electroluminescent diodes, Appl. Phys. Lett. 51 (1988) 913-915.

[38]. P.E. Burrows, Z. Shen, V. Bulovic, D.M. McCarty, S.R. Forrest, J.A. Cronin, M.E. Thompson, Relationship between electroluminescence and current transport in organic heterojunction light-emitting devices, J. Appl. Phys. 79 (1996) 7991-8006.

[39]. L.S. Hung, C.W. Tang, M.G. Mason, Enhanced electron injection in organic electroluminescence devices using an Al/LiF electrode, Appl. Phys. Lett. 70 (1997) 152-154.

[40]. S.R. Cowan, N. Banerji, W.L. Leong, A.J. Heeger, Charge formation, recombination, and sweep-out dynamics in organic solar cells, Adv. Funct. Mater. 22 (2012) 1116-1128.

[41]. C.M. Proctora, M. Kuika, T.-Q. Nguyen, Charge carrier recombination in organic solar cells, Prog. Polym. Sci. 38 (2013) 1941-1960.

[42]. F. Gao, O. Inganäs, Charge generation in polymer–fullerene bulk-heterojunction solar cells. Phys. Chem. Chem. Phys. 16 (2014) 20291-20304.

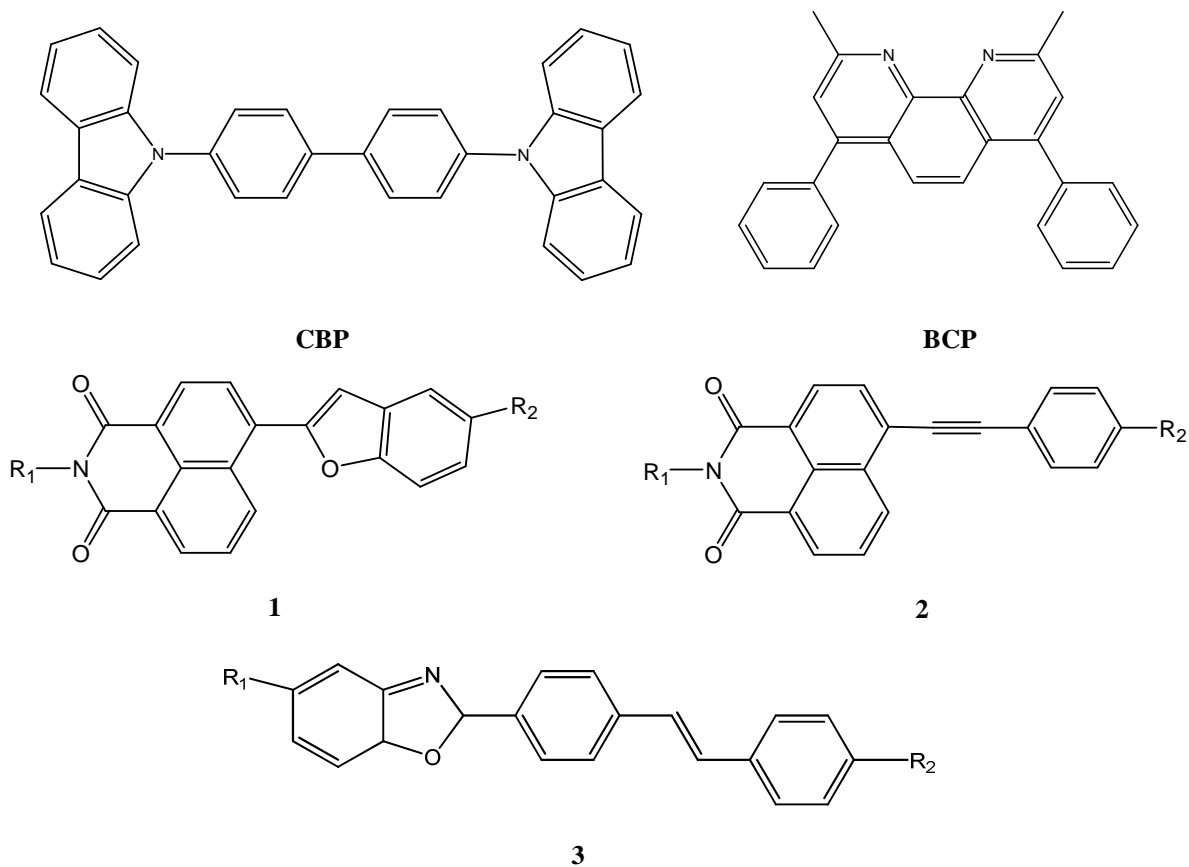

**Figure 1.** Chemical structures of CBP, BCP, **molecule 1**: N-alkyl-4-benzofuran-1,8-naphthalimides, **molecule 2**: N-alkyl-4-arylacetylene-1,8-napthalimides and **molecule 3**: 2-(E)-stilbene benzoxazole. ($R_1$ and $R_2$ indicate side groups in the molecules).

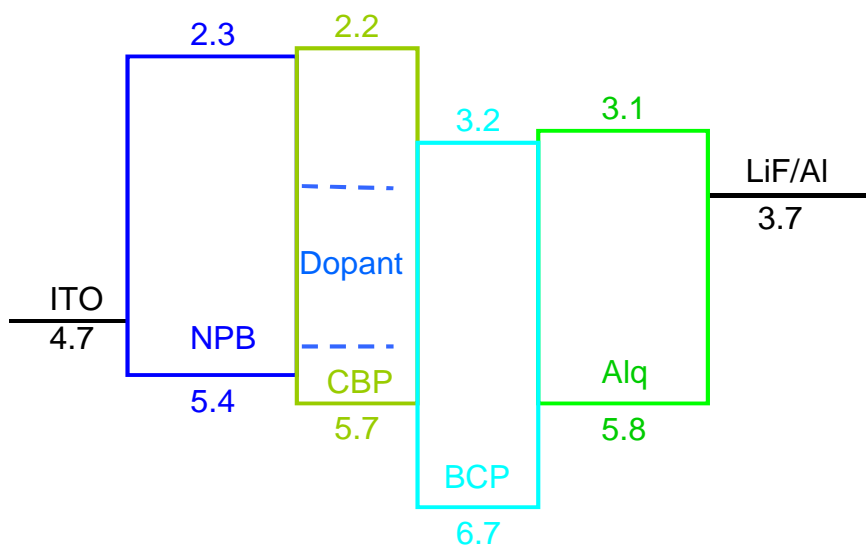

**Figure 2.** Device structure and energy levels for dye molecule doped multilayer OLEDs.

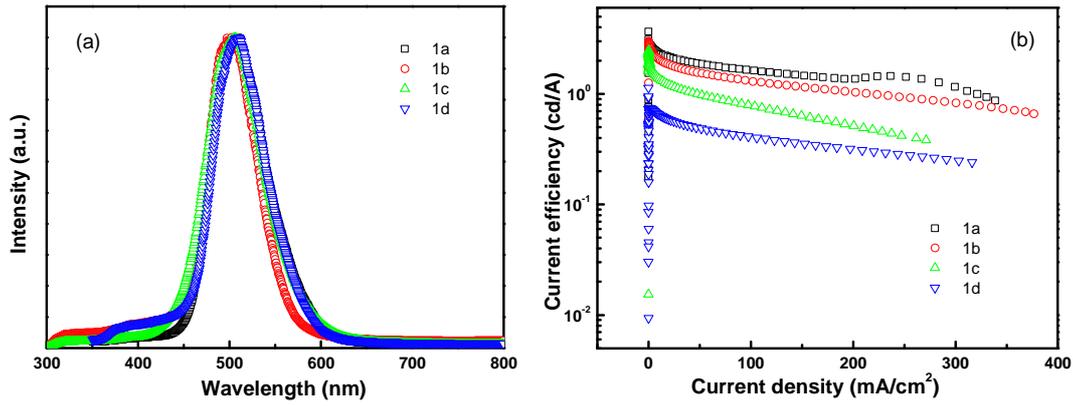

**Figure 3.** Device performance for dye molecule-doped multilayer OLEDs with Molecule 1 doping. (a) EL spectra and (b) Current efficiency-current density.

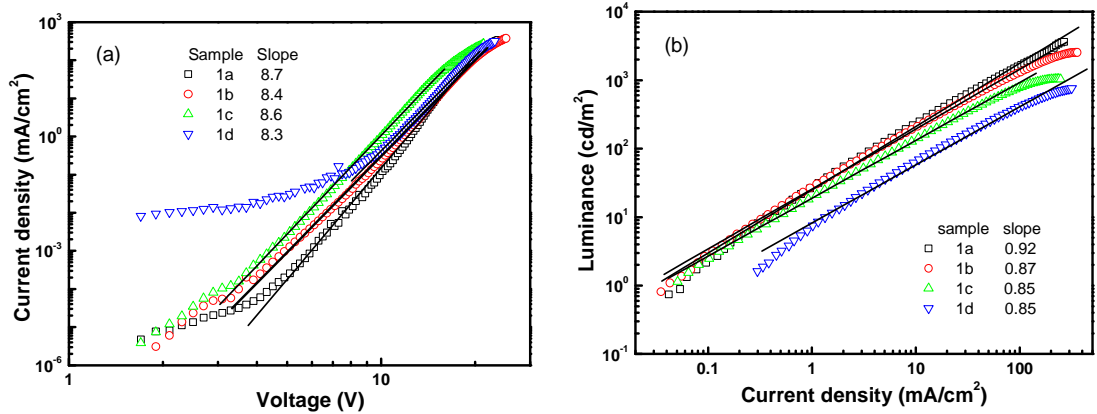

**Figure 4.** Steady state characterization for the devices with Molecule 1 doping. (a) Current density-voltage and (b) Luminance-current density.

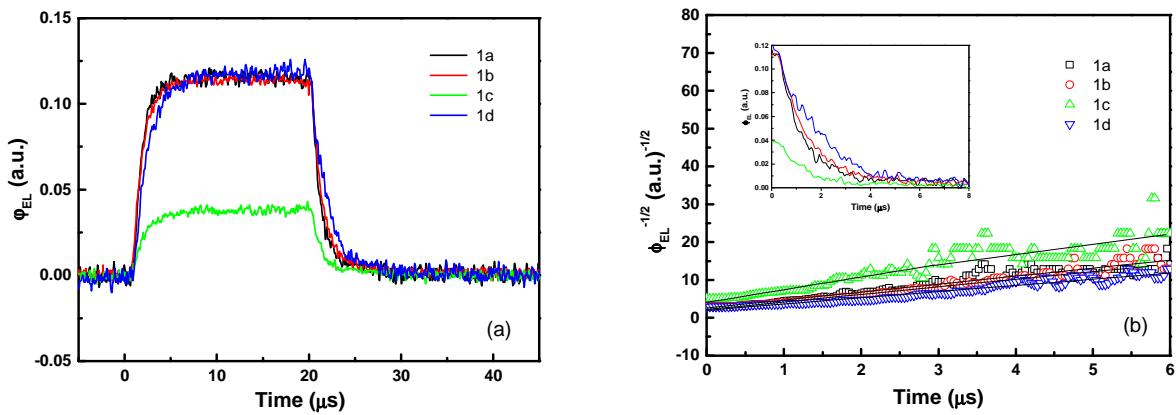

**Figure 5.** Transient EL characterization for the devices with Molecule 1 doping. (a) Transient EL intensity-time and (b) comparison of the transient EL decay at the same equilibrium current density of 170 mA/cm$^2$ plotted in $(\varphi_{EL})^{-1/2}$ versus time scale. The $\varphi_{EL}(t)$ decay curves are shown in the inserted. Here t=0 corresponds to the voltage fall of the pulse.

**Table 1.** List of Molecule 1 with different side groups

| Sample | $R_1$ | $R_2$ | $\lambda_{UV, max}$ (nm) | $\lambda_{FL, max}$ (nm) | Melt point (°C) |
|---|---|---|---|---|---|
| 1a | n-Bu | t-Bu | 397.0 | 508.5 | 170.0-171.5 |
| 1b | n-Bu | $CH_3$ | 398.5 | 508.0 | 140.0-141.0 |
| 1c | n-Bu | Cl | 386.5 | 493.0 | 174.5-176.0 |
| 1d | n-Hexyl | CN | 378.5 | 470.0 | 213.0-214.0 |

**Table 2.** List of device performance and related parameters for Molecule 1

| Sample | Slope (S) | Intercept (A) | γ ($10^{-12}$ $cm^3$/s) | EL efficiency (cd/A) | Luminance (cd/$m^2$) |
|---|---|---|---|---|---|
| 1a | 2.25 | 2.22 | 5.80 | 3.7 | 3640 |
| 1b | 2.03 | 2.11 | 5.22 | 3.0 | 2550 |
| 1c | 3.32 | 4.09 | 3.72 | 2.5 | 1040 |
| 1d | 1.28 | 2.48 | 1.50 | 0.8 | 760 |

* The unit is $10^6$ $(cm^3/s)^{1/2}$ for S, and $(cm^3 s)^{1/2}$ for A.

# Supplementary materials

Experimental results for dye molecule-doped multilayer OLEDs with other molecular systems (2 and 3) are listed as below by use of steady state and transient EL characterization to support the conclusion and discussion in the main text.

**Table s1.** List of small molecules with different side groups

| Sample | $R_1$ | $R_2$ | Melt point (℃) |
|---|---|---|---|
| 2a | n-Bu | H | 146.0-147.0 |
| 2b | n-Bu | $CH_3$ | 170.0-171.0 |
| 2c | n-Bu | F | 145.5-146.5 |
| 2d | n-Hexyl | $CF_3$ | 166.0-167.0 |
| 3a | t-Bu | p-$COOCH_3$ | 222.5-224 |
| 3b | t-Bu | p-$C_6H_6$ | 264-265 |
| 3c | H | 2,5-$OCH_3$ | 140.5-141.5 |

**Table s2.** List of device performance and related parameters

| Sample | Slope (S) | Intercept (A) | $\gamma$ ($10^{-12} cm^3 s^{-1}$) | EL efficiency (cd/A) |
|---|---|---|---|---|
| 2a | 2.22 | 1.66 | 7.80 | 2.6 |
| 2b | 2.08 | 2.09 | 4.53 | 1.9 |
| 2c | 1.97 | 2.16 | 3.63 | 1.2 |
| 2d | 2.67 | 3.04 | 3.36 | 1.0 |
| 3a | 2.76 | 2.40 | 6.34 | 1.4 |
| 3b | 2.04 | 2.43 | 3.39 | 1.2 |
| 3c | 3.42 | 4.71 | 2.54 | 0.8 |

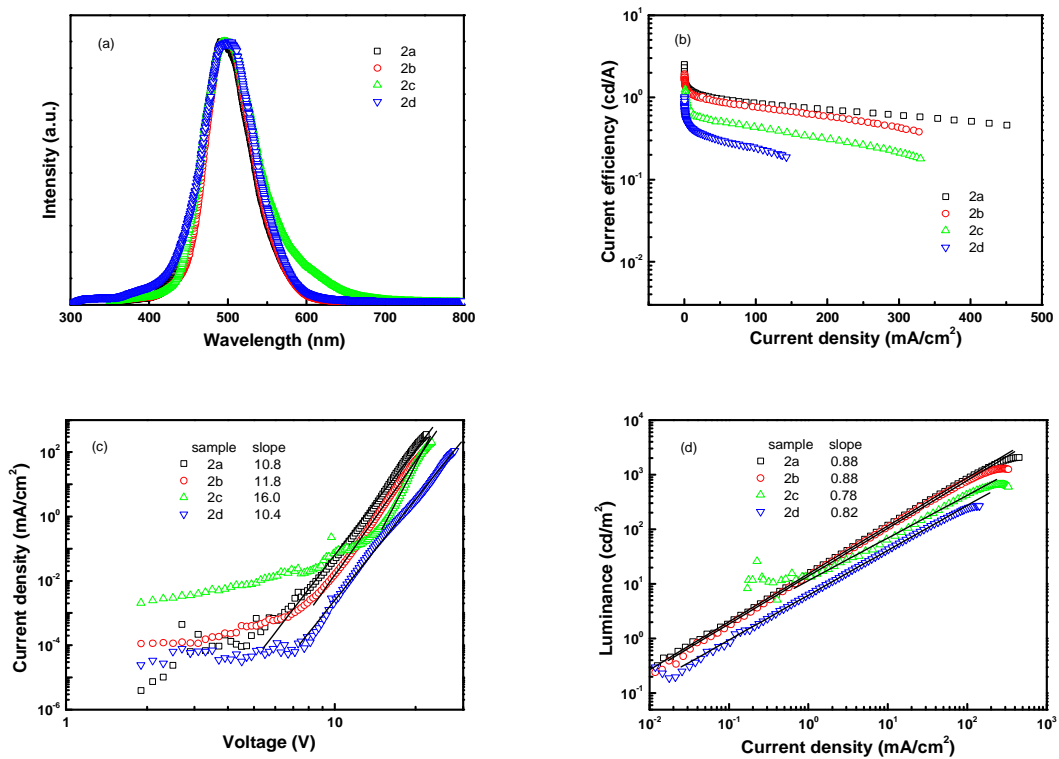

**Figure s1.** Device performance for multilayer OLEDs with molecule 2 doping. (a) EL spectra, (b) Current efficiency-current density, (a) Current density-voltage and (b) Luminance-current density

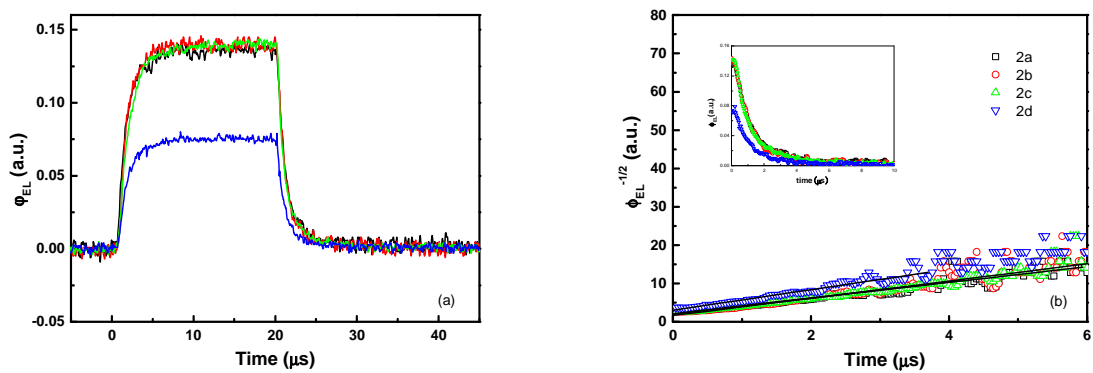

**Figure s2.** Transient EL characterization for multilayer OLEDs with molecule 2 doping. (a) Transient EL intensity-time and (b) comparison of the transient EL decay in dye molecule-doped OLEDs at the same equilibrium current density of 220 mA/cm$^2$ plotted in $(\varphi_{EL})^{-1/2}$ versus time scale. The $\varphi_{EL}(t)$ decay curves are shown in the inserted. Here t=0 corresponds to the voltage fall of the pulse.

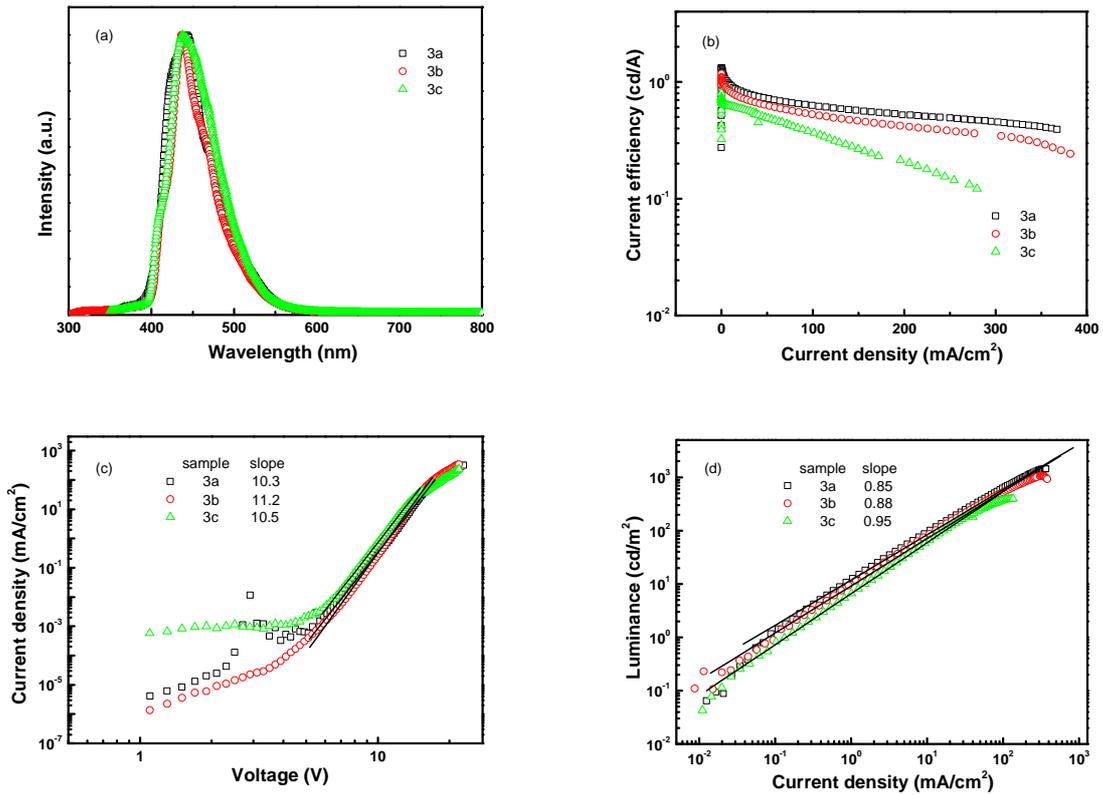

**Figure s3.** Device performance for multilayer OLEDs with molecule 3 doping. (a) EL spectra, (b) Current efficiency-current density, (a) Current density-voltage and (b) Luminance-current density

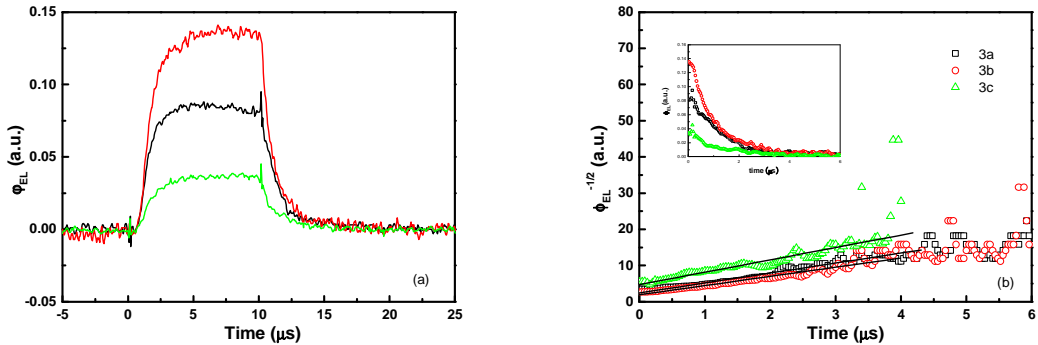

**Figure s4.** Transient EL characterization for multilayer OLEDs with molecule 3 doping. (a) Transient EL intensity-time and (b) comparison of the transient EL decay in dye molecule-doped OLEDs at the same equilibrium current density of 200 mA/cm² plotted in $(\varphi_{EL})^{-1/2}$ versus time scale. The $\varphi_{EL}(t)$ decay curves are shown in the inserted. Here t=0 corresponds to the voltage fall of the pulse.